\documentclass{article}

\usepackage[utf8]{inputenc} 
\usepackage[T1]{fontenc}    
\usepackage[colorlinks=true,linkcolor=black,anchorcolor=black,citecolor=black,filecolor=black,menucolor=black,runcolor=black,urlcolor=black]{hyperref}       
\usepackage{url}            
\usepackage{booktabs}       
\usepackage{amsfonts}       
\usepackage{nicefrac}       
\usepackage{microtype}     
\usepackage{xcolor}         
\usepackage{bm}         
\usepackage{graphicx}  		
\usepackage{amsmath} 
\usepackage{cite} 
\usepackage{diagbox}
\usepackage{subfig}
\usepackage{siunitx}
\usepackage{multirow}
\usepackage{tabularx}
\usepackage{hhline}
\usepackage{arydshln}		
\usepackage{xcolor}
\usepackage{mathtools} 
\usepackage{amsthm}			
\usepackage{amssymb}			
\usepackage{algorithmic}	
\usepackage{algorithm}
\usepackage{pifont}
\usepackage{threeparttable}
\usepackage{fullpage}
            
\usepackage{array}
\usepackage{makecell}
\usepackage{arydshln}
\usepackage[export]{adjustbox}
\usepackage{booktabs}

\usepackage{mathtools} 
\usepackage{microtype} 
\usepackage{multirow}
\usepackage{nicefrac}       
\usepackage{pifont}   

\usepackage{url}
\usepackage{siunitx}
\usepackage{tabularx}
\usepackage{threeparttable}
\usepackage{xcolor}  

\usepackage{dcolumn}
\newcolumntype{d}[1]{D{.}{.}{#1}}
\newcommand\mc[1]{\multicolumn{1}{c}{#1}} 
\newcommand\mcl[1]{\multicolumn{1}{c|}{#1}}

\definecolor{darkblue}{rgb}{0,0 ,0.542}
\definecolor{lightgreen}{rgb}{.9,1,.9}
\definecolor{lightred}{rgb}{1,.415,.415}
\definecolor{lightblue}{rgb}{.415,.415,1}

\newcolumntype{L}[1]{>{\raggedright\arraybackslash}p{#1}}
\newcolumntype{C}[1]{>{\centering\arraybackslash}p{#1}}
\newcolumntype{R}[1]{>{\raggedleft\arraybackslash}p{#1}}

\theoremstyle{plain} 

\def\defn{\,\coloneqq\,}

\def\C{\mathbb{C}}
\def\R{\mathbb{R}}

\def\ebm{{\bm{e}}}

\def\xbm{{\bm{x}}}

\def\ybm{{\bm{y}}}

\def\Abm{{\bm{A}}}

\def\Fbm{{\bm{F}}}

\def\thetabm{{\bm{\theta }}}

\def\Abm{{\bm{A}}}

\def\Fbm{{\bm{F}}}

\def\Mbm{{\bm{M}}}

\def\Dsf{{\mathsf{D}}}

\def\Isf{{\mathsf{I}}}

\def\Tsf{{\mathsf{T}}}

\def\Tsf{{\mathsf{T}}}

\def\Dsf{{\mathsf{D}}}

\def\Isf{{\mathsf{I}}}

\def\xbmast{{\bm{x}^\ast}}
\def\xbmbar{{\overline{\bm{x}}}}

\def\argmin{\mathop{\mathsf{arg\,min}}} 

\title{Overcoming Distribution Shifts in Plug-and-Play Methods with Test-Time Training
}
\date{}

\author{
    Edward P.\ Chandler, Shirin Shoushtari, Jiaming Liu,\\ M.\ Salman Asif, and Ulugbek S.\ Kamilov\\
    \small Washington University in St. Louis, MO 63130, USA\\
    \small University of California Riverside, CA 92521, USA\\
    \small \texttt{\{e.p.chandler, s.shirin,  jiaming.liu, kamilov\}@wustl.edu, \texttt{sasif@ucr.edu} }
}

\begin{document}

\maketitle

\begin{abstract}
\noindent 
Plug-and-Play Priors (PnP) is a well-known class of methods for solving inverse problems in computational imaging. PnP methods combine physical forward models with learned prior models specified as image denoisers. A common issue with the learned models is that of a performance drop when there is a distribution shift between the training and testing data. Test-time training (TTT) was recently proposed as a general strategy for improving the performance of learned models when training and testing data come from different distributions. In this paper, we propose PnP-TTT as a new method for overcoming distribution shifts in PnP. PnP-TTT uses deep equilibrium learning (DEQ) for optimizing a self-supervised loss at the fixed points of PnP iterations. PnP-TTT can be directly applied on a single test sample to improve the generalization of PnP. We show through simulations that given a sufficient number of measurements, PnP-TTT enables the use of image priors trained on natural images for image reconstruction in magnetic resonance imaging (MRI).
\end{abstract}


\section{Introduction}

\medskip\noindent
Many computational imaging problems can be formulated as \emph{inverse problems}, where the goal is to recover an unknown image from a set of noisy measurements. It is common to solve inverse problems by integrating the measurement model characterizing the response of the imaging instrument with a regularizer infusing prior knowledge on the unknown image. There has been considerable recent interest in using deep learning (DL) for designing data-driven image priors~\cite{McCann.etal2017, Lucas.etal2018, Kamilov.etal2023}. DL methods eliminate the need for explicit prior modeling by learning a mapping from measurements to target images using convolutional neural networks (CNN). 

\medskip\noindent
Model-based DL (MBDL) is an extension to traditional DL that integrates the image prior defined through a CNN with the knowledge of the measurement models. For example, plug-and-play priors (PnP) is a well-known MBDL approach that uses pre-trained image denoiser as priors~\cite{Venkatakrishnan.etal2013, Sreehari.etal2016, Kamilov.etal2023}. Other MBDL widely-used MBDL approaches include deep unfolding (DU) and deep equilibrium (DEQ) learning, both of which rely on the integration of the measurement model during the training of the image prior~\cite{Hosseini.etal2019, Ongie.etal2020, Monga.etal2021, Gilton.etal2021, Liu.etal2022a}. While both DU and DEQ interpret iterations of image reconstruction as neural network layers, the memory complexity of DEQ is independent of the number of unfolded iterations.

\medskip\noindent
Much of the existing research on MBDL has focused on the scenarios where the statistical distribution of the training data matches that of the testing data. While this strategy has led to significant theoretical and algorithmic innovations, it does not address the issue of the performance gap due to data distribution shifts. For example, image priors trained with a specific distribution in PnP, performs poorly on samples from different distributions~\cite{shoushtari2022}. Thus, distribution shifts limit the applicability of priors pre-trained for one class to another one.

\medskip\noindent
Domain adaptation refers to a class of DL techniques for improving the performance of a learned model on a target task containing insufficient annotated data by using the knowledge learned by the model from another related task with adequate labeled data~\cite{Venkateswara.etal2017, Farahani.etal2021}. 
\emph{Test-time training (TTT)} was recently proposed as a domain adaptation strategy based on self-supervised optimization of the learned model utilizing only test-time measurements~\cite{sun2020}. The TTT strategy was also recently used in the context of imaging inverse problems to address domain shifts in end-to-end image reconstruction with DL for accelerated magnetic resonance imaging (MRI)~\cite{darestani2022}.

\medskip\noindent
In this paper, we investigate TTT in the context of PnP methods. We propose \emph{PnP-TTT} as a method for overcoming the performance gap in PnP due to data distribution shifts. PnP-TTT uses DEQ to update the weights of the CNN prior in PnP at test-time. The DEQ learning in PnP-TTT is used to minimize a self-supervised loss at the fixed points of PnP iterations for one test sample. We also present numerical results showing that DEQ training in PnP-TTT can significantly boost the performance of the shifted priors. We evaluate the proposed method on image reconstruction for compressed sensing MRI (CS-MRI), where we recover MRI images from subsampled Fourier measurements. Our results show that given enough measurements, PnP-TTT can close the gap due to distribution shift between test and training data. 
It is worth mentioning that our method can also be applied to other tasks and different variants of PnP, highlighting its broader applicability for inverse problems in computational imaging. 



\begin{algorithm}[!tp]
\caption{Test-Time Training for Plug-and-Play Methods}
     \begin{algorithmic}
     \renewcommand{\algorithmicrequire}{\textbf{input:}}
     \REQUIRE forward model $\mathbf{A}$, PnP initialization $\boldsymbol{x}_0$, measurement $\boldsymbol{y}$, denoiser $\Dsf_{\theta}$, and  \texttt{numIter}$\geq0$
        \STATE$\boldsymbol{x}^*_0= \texttt{PnP}(\boldsymbol{x}_0, \boldsymbol{y}; \Dsf_{\theta})$
        \FOR {$i = 1$ to \texttt{numIter}}
        \STATE $l= \texttt{LOSS}(\mathbf{A}\boldsymbol{x}^*_{i-1}, \boldsymbol{y})$
        \STATE $\texttt{DEQ\_GRAD}(l, \theta)$ \qquad \COMMENT{Update parameters $\theta$ using DEQ}
        \STATE$\boldsymbol{x}^*_i= \texttt{PnP}(\boldsymbol{x}_0, \boldsymbol{y}; \Dsf_{\theta})$
        \ENDFOR
     \RETURN $\boldsymbol{x}^*_i$ 
    \label{alg: proposed}
     \end{algorithmic}
 \end{algorithm}
 
\section{Background}

\subsection{Inverse Problems}
\medskip\noindent
We consider the problem of  recovering an image $\xbm \in \C^n $ from its noisy measurement $\ybm = \Abm \xbm + \ebm$, 
where $\Abm\in \C^{m\times n}$ is the \emph{measurement operator} and $\ebm \in \C^m$ is additive white Gaussian noise (AWGN).
We can formulate the problem as a regularized optimization problem
\begin{equation}
    \xbmast = \argmin_{\xbm\in \R^n} f(\xbm) \quad \text{with} \quad  f(\xbm) =  g(\xbm) + h(\xbm), \label{eq: optimization problem}
\end{equation}
where $g$ is the \emph{data-fidelity} term used to ensure the consistency of the solution with the measurement and $h$ is the \emph{regularization} term that infuses prior knowledge. For example, the least-squares loss is a widely-used data-fidelity term $g(\xbm) = \frac{1}{2}\|\ybm - \Abm \xbm\|_2^2$ and total variation (TV) is commonly used as the regularizer~\cite{beck2009}. 

\subsection{Plug-and-Play Priors}
\medskip\noindent
PnP framework includes a family of methods that incorporate the measurement model with CNN denoisers to solve inverse problems~\cite{Kamilov.etal2023}. PnP methods can be interpreted as a fixed-point iteration of some high-dimensional operator where the CNN takes the role of the prior. For example, the
\emph{proximal gradient method (PGM)} variant of PnP can be expressed as 
\begin{align}
    \label{eq: PnP}
    &\xbm^k = \Tsf_{\thetabm}(\xbm^{k-1}) \quad \text{with} \quad \Tsf_{\thetabm} \defn \Dsf_{\thetabm}( \Isf- \gamma \nabla g), 
\end{align}
where $\Dsf_\thetabm$ is the denoiser, $g$ is the data-fidelity term, $\nabla g$ is the gradient of $g$, $\Isf$ is the identity mapping, and $\gamma>0$ is the step-size. The PnP method in~\eqref{eq: PnP} is commonly refered to as PnP-PGM.

\begin{figure*}[tp]
 \centering
 \includegraphics[scale=.91]{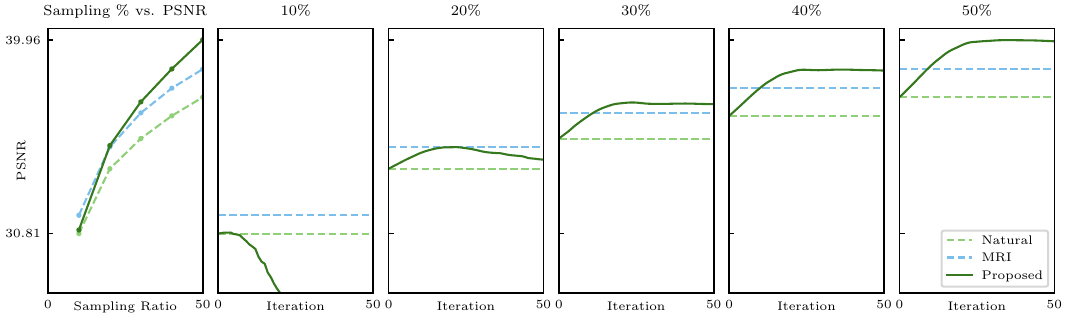} 
 \caption{Evaluation of PnP-TTT for different sampling ratios in accelerated MRI. The left-most chart displays the best PSNR performance achieved by PnP-TTT at different sampling ratios. 
 The remaining five charts show PSNR over all examples at each TTT iteration. Note that the best performance is above the lower baseline for all the sampling ratios; however, TTT eventually overfits to the test-time measurement, reducing performance. Additionally, note that at larger sampling ratios, the performance of PnP-TTT prior can surpass that of the matched prior due to the DEQ training.}
\label{fig: CS vs TTT}
\end{figure*}

\begin{table*}[tp]
\caption{PSNR (dB) values for accelerated MRI with matched, mismatched, and PnP-TTT priors.}
\begin{center}
\renewcommand\arraystretch{1.2}
\setlength{\tabcolsep}{3pt}
\begin{tabular}{c 
                d{2.2}
                d{2.4} | 
                d{2.2}
                d{2.4} |
                d{2.2}
                d{2.4} |
                d{2.2}
                d{2.4} |
                d{2.2}
                d{2.4}}
    \hline
    Radial CS Ratio & \multicolumn{2}{c}{$10\%$} & \multicolumn{2}{c}{$20\%$} & \multicolumn{2}{c}{$30\%$} & \multicolumn{2}{c}{$40\%$} & \multicolumn{2}{c}{$50\%$}\\
    \cline{2-11} 
     & \mc{PSNR} & \mcl{SSIM} & \mc{PSNR} & \mcl{SSIM} & \mc{PSNR} & \mcl{SSIM} & \mc{PSNR} & \mcl{SSIM} & \mc{PSNR} & \mc{SSIM} \\
    \Xhline{4\arrayrulewidth}
    Natural prior & 30.81  &  0.9351 &  33.87  &  0.9648 &  35.29  &  0.9721 &  36.37  &  0.9759 &  37.25 &  0.9782  \\
    MRI Prior & 31.67  &  0.9468 &  34.9  &  0.9707 &  36.51  &  0.9773 &  37.67  &  0.9807 &  38.57 &  0.9828  \\
    PnP-TTT (Ours) & 30.97  &  0.938 &  34.97  &  0.9718 &  37.03  &  0.9796 &  38.58  &  0.984 &  39.96 &  0.9873 \\
    \hline
    PnP-TTT $-$ Natural & 0.16  &  0.0029 &  1.1  &  0.007 &  1.74  &  0.0075 &  2.21  &  0.0081 &  2.71 &  0.0091\\
    \hline
\end{tabular}
\label{tab: performance}
\end{center}
\end{table*}

\subsection{Deep Equilibrium Models}
\medskip\noindent
DEQ is a recent approach for training MBDL architectures in a memory-efficient way~\cite{Gilton.etal2021}. DEQ uses implicit differentiation for training possibly infinite-depth networks by backpropagating through the fixed points of an operator. For the operator defined in eq.~\eqref{eq: PnP}, the output of the DEQ is implicitly expressed as 
\begin{equation}
    \label{eq:DeqFixedPoint}
    \xbmbar = \Tsf_\thetabm (\xbmbar),
\end{equation}
where $\Tsf_\thetabm$ is the operator parameterized by $\thetabm$, and $\xbmbar$ is the fixed point acquired using fixed point iterations in the forward pass of DEQ. The connection of DEQ and PnP has inspired end-to-end training of CNN denoisers as model dependant priors in many imaging problems such as MRI~\cite{Gilton.etal2021} and computed tomography (CT)~\cite{Liu.etal2022a}.

\medskip\noindent
The prior $\Dsf_\thetabm$ in DEQ is trained by minimizing the loss between the fixed points from eq.~\eqref{eq:DeqFixedPoint} and the ground truth image $\xbmast$
\begin{equation}
    \label{eq:LossDeq}
    \ell(\thetabm)= \frac{1}{2}\|\Tsf_{\theta}(\xbmbar)-\xbmast\|_2^2.
\end{equation}
Implicit differentiation of the fixed points yields the gradient of the loss with respect to $\thetabm$ in the backward pass as the following
\begin{equation}
    \label{eq:GradDeqLoss}
    \nabla\ell(\thetabm)= (\nabla_\thetabm \Tsf_\thetabm(\xbmbar))^\Tsf \left(\Isf - \nabla_\xbm \Tsf_{\thetabm}(\xbmbar)\right)^{-\Tsf}(\xbmbar-\xbmast),
\end{equation}
where $\Isf$ is the identity mapping and $\ell$ is the loss in eq.~\eqref{eq:GradDeqLoss}. 

\subsection{Test-Time Training}

\medskip\noindent
Current PnP methods are built on the premise that the prior represents the same distribution as that of the desired solution. However, it is common to observe distribution shifts between training and testing data. In some scenarios, there are insufficient training samples to train a DL network as the prior, and as a result, alternative priors trained on a shifted distribution are used. The distribution shift results in suboptimal reconstruction performance. TTT has been proposed to reduce the performance gap due to distribution shift in various tasks~\cite{sun2020}. The key idea of TTT is to update the shifted model's weight at test-time by minimizing  a self-supervised loss 
\begin{equation}
    \thetabm^\ast = \argmin_{\thetabm} \ell_{\textsf{\tiny sup}} \left( \Dsf_{\thetabm}(\ybm), \ybm \right), 
\end{equation}
where $\Dsf_\thetabm$ is the neural network and $\ybm$ is a test sample. 
Depending on the selection of $\ell_{\textsf{\tiny sup}}$, TTT has shown improved performance in many imaging tasks. For example, it can be used to improve the MRI reconstruction using DL models trained in an end-to-end matter on shifted distributions~\cite{darestani2022}. 
In this scenario, the self-supervised loss proposed is the following
\begin{equation}
    \label{eq:SSlossTTT}
    \ell_{\textsf{\tiny sup}}(\thetabm) = \frac{\left\Vert \Abm \Dsf_{\thetabm}\left(\Abm^{\dagger} \ybm\right) - \ybm  \right\Vert_1}{\Vert \ybm \Vert_1}, 
\end{equation}
where $\Abm$ is the measurement model, $\Abm ^\dagger$ is the Hermitian transpose, and $\ybm$ is the test-time measurement. Note that as opposed to \eqref{eq:LossDeq}, TTT in~\eqref{eq:SSlossTTT} does not need ground truth reconstruction to compute $\ell_{\textsf{\tiny sup}}$. Losses other than the normalized $\ell_1$ norm can work as well~\cite{darestani2022}.

\subsection{Our contribution}
\medskip\noindent
We propose PnP-TTT as a novel approach for enhancing  the performance of image reconstruction for PnP methods under distribution shifts. Our approach involves domain adaptation for a shifted pre-trained image prior through TTT using DEQ to close the distribution gap, which only requires a test single measurement. Our results show that PnP-TTT can improve the performance significantly given sufficient measurement for shifted priors with minimal computational cost. 

\section{Method}
\medskip\noindent
We now present our method for domain adaption of image priors in PnP. We consider the PnP-PGM algorithm in eq.~\eqref{eq: PnP} and run it until its convergence. In practice, we find that about $100$ iterations of PnP-PGM are sufficient in our configuration. We can update the weights of the image prior on a test measurement by minimizing the following self-supervised loss
\begin{equation}
    \label{eq:PnPTTTloss}
    \ell_{\textsf{\tiny sup}}(\thetabm) = \left\Vert  \Abm \Tsf_\thetabm(\xbmbar) - \ybm \right\Vert_2^2, 
\end{equation}
where $\xbmbar$ is the fixed-point of PnP-PGM defined in eq.~\eqref{eq:DeqFixedPoint} and $\Tsf_{\theta}$ is the operator defined in \eqref{eq: PnP}. We use the DEQ to compute the gradient of $\ell_{\textsf{\tiny sup}}$ at test-time using implicit differentiation.  

\medskip\noindent
We follow a method similar to~\cite{zhang2017, miyato2018} to train the image priors using the DnCNN architecture, with batch normalization layers replaced with spectral normalization to control the Lipschitz constant of the denoisers. DnCNN is trained as a denoiser for AWGN level $\sigma=5$. During the training stage we do not use DU or DEQ so that the learned prior model is purely an image denoiser. We use 400 CBSD to train natural prior on grayscale images of size $180\times180$~\cite{Martin.etal2001}. We train MRI priors on MRI brain images of size $256\times256$~\cite{zhang2018}. 

\medskip\noindent
For test-time training, we initialize PnP-PGM with $\xbm^0 = \bm{0}$ and $100$ iterations in the forward pass of DEQ, using the denoising prior acquired as described above. We use Nesterov acceleration~\cite{nesterov2003}, and set stepsize $\gamma = 1$. We use $100$ iterations and Anderson acceleration in the backward pass of DEQ~\cite{anderson1965}. We allow TTT to run for 50 iterations, using SGD to update the parameters $\theta$ with a step size of $1\times10^{-5}$. At inference, using the adapted prior, we again run for $100$ iterations with Nesterov acceleration in PnP-PGM with step size $\gamma = 1$. 
Note that once all $50$ TTT iterations are performed for a particular measurement, $\theta$ is reset to the non-domain-adapted weights. Since the goal of PnP-TTT is to overcome the performance gap from a distribution shift between train- and test-time, performing TTT on as many measurements are available at test-time may be beneficial. Future experiments could examine if there is any performance improvement when using multiple measurements instead of only one for PnP-TTT.

\medskip\noindent
The measurement model for a single-coil, accelerated MRI with radial Fourier sampling can be modeled as  $\Abm = \Mbm \Fbm$, where $\Mbm$ is the diagonal sampling matrix and $\Fbm$ is the Fourier transform. We investigate five different sampling ratios $(m/n)$ in the experiments. For the experiments reported here, we consider noiseless scenario; however, we expect similar performance of PnP-TTT under moderate amoungs of noise.

\begin{figure*}[!tp]
 \centering
 \includegraphics[scale=0.91]{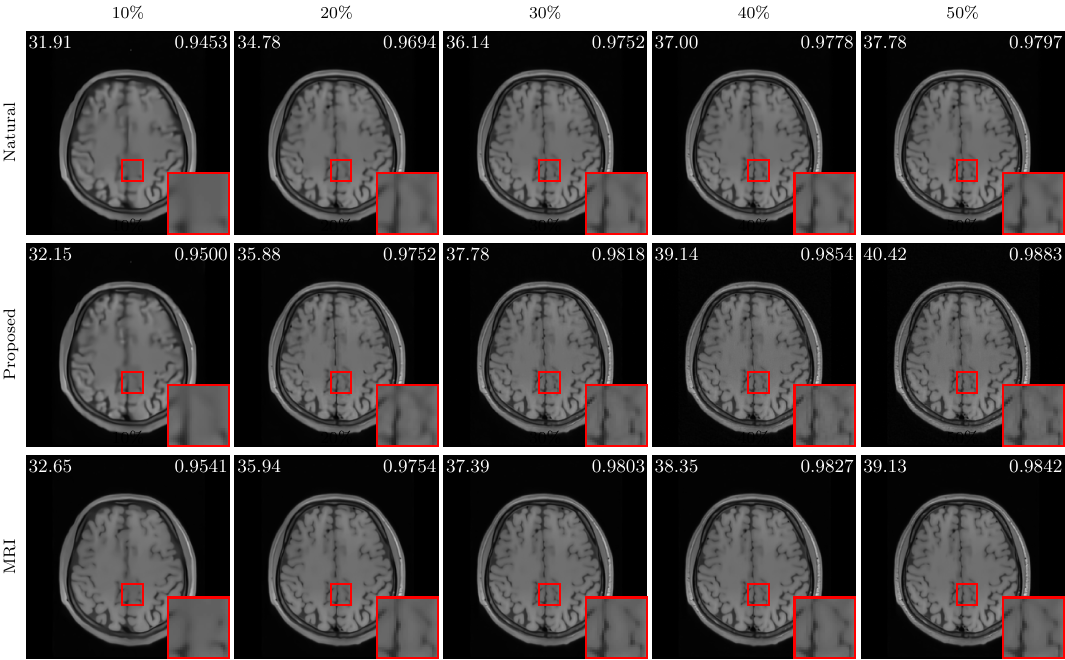} 
 \caption{Visual evaluation of various priors at different CS ratios for CS-MRI problem with corresponding PSNR (upper left) and SSIM (upper right). Note that natural priors (top row) performs suboptimally compared to matched MRI prior (bottom row). Additionally, note the improvement due to the usage of PnP-TTT framework for all CS ratios (middle row).}
\label{fig: recon img}
\end{figure*}

\section{Results}
\medskip\noindent
We test our proposed method by reconstructing ten brain MRI images selected from the test dataset of \cite{zhang2018} with mismatched DnCNN prior trained on natural images. Due to distribution shift, natural priors demonstrate suboptimal performance for the MRI task. To establish a performance baseline, we compare the result of the proposed method with those obtained by mismatched natural prior and matched MRI prior. Specifically, we consider the performance achieved by natural prior as the lower baseline, and that achieved by the MRI prior as the upper baseline. Our proposed PnP-TTT seeks to enhance the performance of a mismatched natural prior so as to approach that of a matched MRI prior. 

\medskip\noindent
Table \ref{tab: performance} reports the best results achieved for five CS ratios: $10,\ 20,\ 30,\ 40,\ \text{and}\ 50$. It can be seen that PnP-TTT can close the performance gap for CS ratios of $20$ and more, while for a CS ratio of $10$, it can make an improvement compared to the lower baseline (mismatched natural prior). The reconstruction quality is quantified using peak signal-to-noise ratio (PSNR) in dB and the structural similarity index measure (SSIM).

\medskip\noindent
Figure \ref{fig: CS vs TTT} illustrates two empirical results: (\emph{a}) The empirical performance of PnP-TTT at testing for different CS ratios (\emph{left figure}), (\emph{b}) The empirical performance during test-time training (\emph{remaining five figures}). Note that during test-time training, the prior can overfit to the measurement. Thus, in practice it is necessary to hold out some measurements to use early stopping during TTT~\cite{darestani2022}, although we have not included such results in this paper. 
The visual results can be found in Figure \ref{fig: recon img}. It can be seen both empirically and visually that PnP-TTT can shorten the gap due to distribution shift, close it completely, or go beyond closing it given the CS ratios.

\section{Conclusion}
\medskip\noindent
We present PnP-TTT as a novel framework for closing the performance gap that arises due to mismatched priors in imaging inverse problems solved by PnP framework. PnP-TTT achieves this by adapting the mismatched priors during the testing phase by using DEQ training to update  the weights of the mismatched priors. One of the main advantage of PnP-TTT is that one can use mismatched priors on a shifted distribution without the need to do additional training on other samples from the shifted distribution. Instead, the prior can simply be adapted to the test-time measurements. Our results show that PnP-TTT can  significantly enhance the performance, achieving performance comparable to that of using a matched prior during inference. Furthermore, this work demonstrate that priors from different tasks can be used interchangeably in scenarios with shifted distribution without the loss of performance. 

\section*{Acknowledgement}
This material is partially based upon work supported by the NSF CAREER award under grant CCF-2043134 and by the Gordon and Betty Moore Foundation grant 11396.

{\small
\bibliographystyle{IEEEbib}
\bibliography{references}
}

\end{document}